# A Universal Mirror-stacking Approach for Constructing Topological Bound States in the Continuum

Luohong Liu,[1] Tianzi Li,[1] Qicheng Zhang,[1] Meng Xiao,[1,2] and Chunyin Qiu[1*]

[1]Key Laboratory of Artificial Micro- and Nano-Structures of Ministry of Education and
School of Physics and Technology, Wuhan University, Wuhan 430072, China

[2]Wuhan Institute of Quantum Technology, Wuhan, 430206, China

[*] To whom correspondence should be addressed: cyqiu@whu.edu.cn

*Abstract.* Bound states in the continuum (BICs) are counter-intuitive localized states with eigenvalues embedded in the continuum of extended states. Recently, nontrivial band topology is exploited to enrich the BIC physics, resulted in topological BICs (TBICs) with extraordinary robustness against perturbations or disorders. Here, we propose a simple but universal mirror-stacking approach to turn nontrivial bound states of any topological monolayer model into TBICs. Physically, the mirror-stacked bilayer Hamiltonian can be decoupled into two independent subspaces of opposite mirror parities, each of which directly inherits the energy spectrum information and band topology of the original monolayer. By tuning the interlayer couplings, the topological bound state of one subspace can move into and out of the continuum of the other subspace continuously without hybridization. As representative examples, we construct one-dimensional first-order and two-dimensional higher-order TBICs, and demonstrate them unambiguously by acoustic experiments. Our findings will expand the research implications of both topological materials and BICs.

*Introduction.*—Bound states in the continuum (BICs), which are spatially confined modes coexisting with a continuous spectrum of extended states, are first proposed by von Neumann and Wigner in 1929 [1,2]. Since the initial proposal in quantum mechanics, BICs have been unveiled in electromagnetic [3-23], acoustic [24-31], and water [27,32] wave systems with a variety of physical mechanisms [2], such as symmetry incompatibility [5,24-26,33], separations of coordinate variables [34-36], parameter tuning [3,4,6,7,28,29,31,32,37], and inverse construction [32,38-40]. Because BICs defy conventional wisdom of confining waves, their realization in different systems will definitely provide surprises and advances in fundamental physics. Extensive applications have been proposed for BICs due to their unique properties (e.g. strong localization and tunable high-Q factor), such as in designing narrow-band filters, biological and chemical sensors, and low-threshold lasers [2,13,18,41-43].

On the other hand, topological materials have attracted tremendous attention over the past decades, ranging from condensed matter physics [44-46] to photonics [47,48], acoustics [49,50], and so on [51]. One fundamental feature of such fascinating phases is the bulk implication of the symmetry-protected boundary modes. Naturally, a combination of the topological band theory and the BIC physics triggers the concept of topological BICs (TBICs) [52-62], which are symmetry-protected and cannot be removed except by large parametric variations. Not only does the emergence of TBICs expand the scope of the established bulk-boundary correspondence (given the fact that boundary-localized states do not hybridize with the bulk surrounding even in the absence of a band gap), but also, more importantly, the topological nature endows the TBICs with inherent protection and thus significantly enhances their ability of controlling waves. To the best of our knowledge, however, there are few studies (especially experimental ones) on TBICs, comparing with the widely-explored topological materials and BIC physics. Particularly, the existed TBIC models are case to case [54-58] or strongly rely on insight [53,59]. Therefore, finding an easily-generalizable TBIC design route is of great significance beyond doubt.

In this Letter, we propose a simple but universal mirror-stacking approach to achieve TBICs. In contrast to most of the TBIC models [54-58] where some spatial symmetries of protecting band topology must be preserved to keep BICs, here the symmetries that induce the band topology and BICs are independent with each other. As sketched in Fig. 1, we stack a pair of identical monolayers with mirror symmetry, each of which features a topological bound state at its boundary [Fig. 1(a)]. The presence of the mirror symmetry enables a classification of the states into two subspaces according to their mirror parities. By tuning the interlayer couplings, the topological bound states can energetically move into and out of the bulk continuums of opposite parities without hybridization, which gives rise to TBICs at appropriate interlayer coupling [Fig. 1(b)]. We examine this idea with mirror-stacked one-dimensional (1D) Su-Schrieffer-Heeger (SSH) model and two-dimensional (2D) quadrupole model. Experimentally, by using acoustic metamaterials consisting of coupled cavity resonators, we observe the associated first-order and higher-order TBICs, which feature zero-dimensional (0D) topological boundary states in the 1D and 2D bulk continuums, respectively.

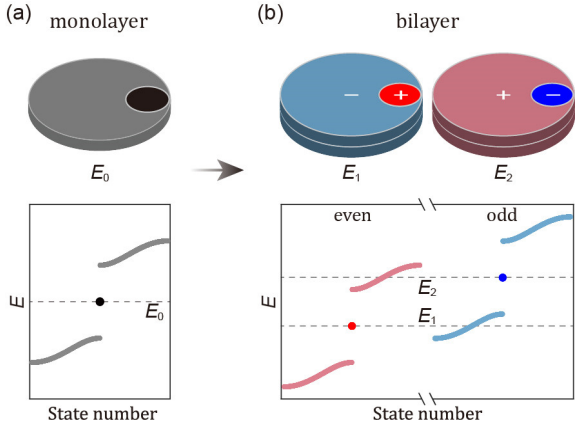

FIG. 1. Mirror-stacking approach for constructing TBICs. (a) Top: Schematic of a monolayer system with a topological bound state at its boundary. Bottom: The associated energy spectrum, which illustrates a nontrivial boundary state (dark) spectrally isolated from the bulk continuum (gray). (b) Top: Mirror-stacked bilayer system that supports two different TBICs. Bottom: Bilayer energy spectrum divided into two independent subspaces according to their mirror parities. At appropriate interlayer coupling, the boundary states can energetically coexist with the bulk ones of opposite parities. Throughout this Letter, we use red and blue to characterize the even (+) and odd (−) parities of the states, respectively.

*Tight-binding models of two concrete examples.*—As depicted in Fig. 2(a), we consider first two directly-coupled 1D SSH chains. The model Hamiltonian reads $H = \tau_0 \otimes h - t_c \tau_1 \otimes I$, where $h = -(t_0 + t_1 \cos k_x)\sigma_1 - t_1 \sin k_x \sigma_2$ represents that of a single chain, $I$ is an identity matrix of the same order of $h$, $-t_0$ and $-t_1$ are two dimerized intralayer couplings, $-t_c$ is the interlayer coupling, and $\sigma$ and $\tau$ are Pauli matrices acting on the intra- and interlayer sublattices, respectively. The bilayer system satisfies the mirror symmetry $[M_z, H] = 0$ with mirror operator $M_z = \tau_1 \otimes I$. After a simple similarity transformation $\widetilde{M}_z = S^{-1} M_z S$ and $\widetilde{H} = S^{-1} H S$, where the transformation matrix $S = \tau_0 \otimes I - i\tau_2 \otimes I$ [63], the mirror operator can be diagonalized into $\widetilde{M}_z = I \oplus (-I)$, and meanwhile, the bilayer Hamiltonian can be decoupled into two independent subspaces according to their mirror parities, i.e., $\widetilde{H} = h_\text{even} \oplus h_\text{odd}$. More details can be seen in Supplemental Material (SM) [64]. Exactly, the Hamiltonian components $h_\text{even} = h - t_c I$ and $h_\text{odd} = h + t_c I$ correspond to two monolayer SSH models of opposite onsite energies $\mp t_c$ [Fig. 2(a)]. This results in pairwise split bulk bands according to their mirror parities [Fig. 2(b)], and more importantly, each copy inherits the original monolayer band topology, which is characterized by a quantized dipole moment in the presence of inversion symmetry. For any nontrivial phase with $t_1/t_0 > 1$, the topological 0D edge state can move continuously into and out of the 1D bulk continuum of opposite parity by tuning the interlayer coupling $t_c$, which contributes a TBIC phase in the phase diagram [Fig. 2(c)]. The formation of the (first-order) TBIC can be visualized more clearly from the energy spectra of finite-sized systems plotted for a fixed intralayer coupling ($t_1/t_0$) but varied interlayer couplings ($t_c/t_0$) [Fig. 2(d)]. Hybridization is forbidden between the energetically degenerate bound and continuum states, since they belong to the subspaces of different parties. It is worth emphasizing that the TBIC is topologically robust against any perturbation that respects the mirror symmetry (see SM [64]).

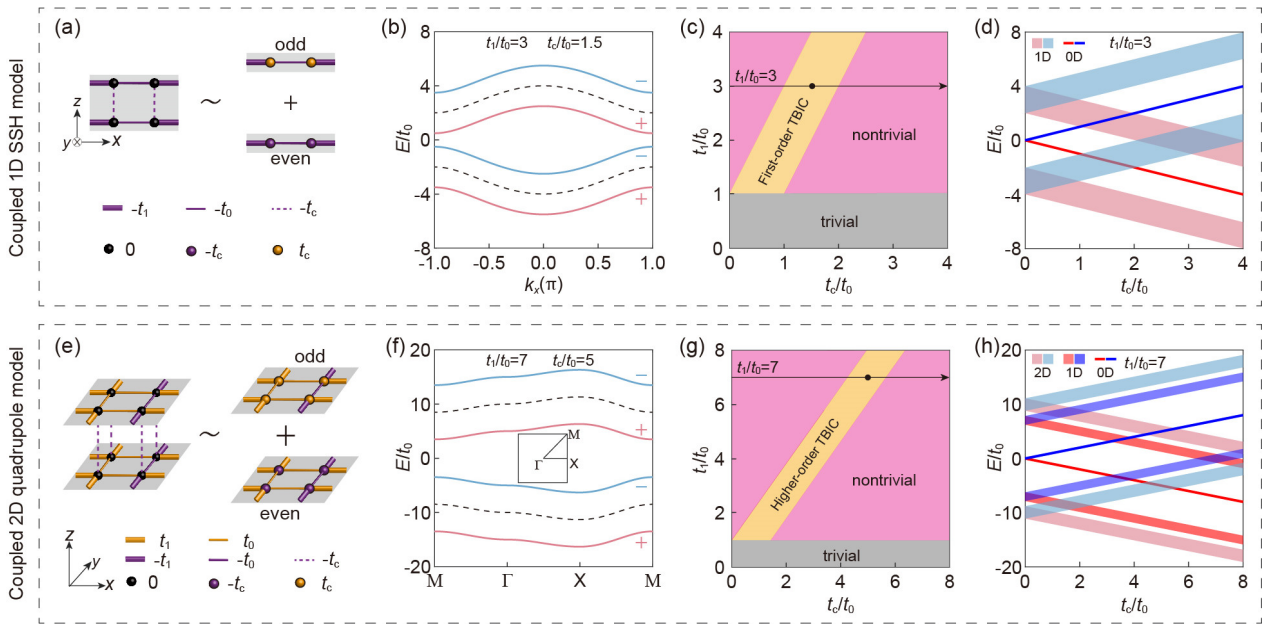

FIG. 2. Model examples for constructing 1D first-order and 2D higher-order TBICs. (a) Unit cell of the directly-coupled 1D SSH model and its effective decomposition (assuming that $t_0, t_1, t_c > 0$). (b) Bilayer band structure (color solid lines) exemplified with specific intralayer and interlayer hoppings, together with its monolayer counterpart (black dashed lines) for comparison. (c) Phase diagram derived for the 1D system. (d) Energy spectra for the finite-sized systems with a fixed intralayer hopping but different interlayer hoppings. It shows that 0D edge states can coexist with 1D bulk states of opposite parity. (e)-(h): Similar to (a)-(d), respectively, but for the directly-coupled 2D quadrupole model, which demonstrate the presence of the higher-order TBICs (featuring 0D corner-localized states embedded in 2D bulk states). Note that each band in (f) is twofold degenerate.

Our mirror-stacking approach can be extended to any monolayer topological system, where $h$ is replaced by the corresponding monolayer Hamiltonian. This is exemplified by the mirror-stacked quadrupole model [Figs. 2(e)-2(h)], where $h = (t_0 + t_1 \cos k_x)\rho_0\sigma_1 + t_1 \sin k_x \rho_0 \sigma_2 + (t_0 + t_1 \cos k_y)\rho_1\sigma_3 + t_1 \sin k_y \rho_2 \sigma_3$ [66,67], with $\sigma$ and $\rho$ being Pauli matrices acting on the $x$- and $y$-directed sublattices, respectively. For the case of $t_1/t_0 > 1$, the monolayer quadrupole model exhibits nontrivial higher-order band topology (characterized by a quantized quadrupole moment in the presence of reflection symmetries), manifested as symmetry-protected bound states at the sample corners. This enables a high-order TBIC phase [Fig. 2(g)] that features energetically degenerate 0D corner states and 2D bulk states of opposite parities [Fig. 2(h)], once introducing the interlayer coupling $t_c$. Again, the presence of odd and even mirror subspaces refrains from the hybridization between the bound and continuum states. Note that there are extra BIC phases in this system due to the presence of the 1D trivial edge states, manifested as 0D corner states embedded in 1D edge states or 1D edge states embedded in 2D bulk states (see SM [64]). The coexistence of multiscale BICs, which could be of great interest, will be systematically explored in the future.

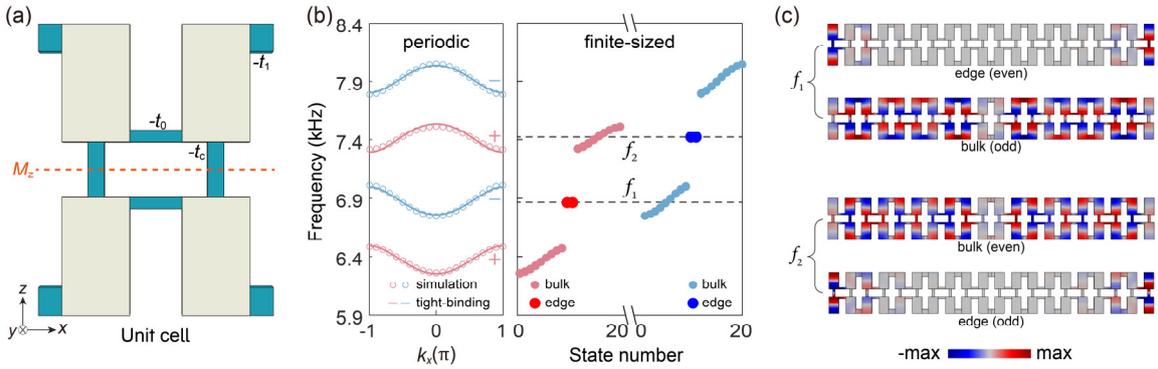

FIG. 3. Acoustic emulations of the mirror-stacked bilayer SSH model. (a) Unit cell consisting of air cavities (white) and narrow tubes (cyan). (b) Left: Band structure (circles) simulated for the cavity-tube structure, matching well that of the tight-binding model (lines). Right: Energy spectra simulated for a finite system of 10 unit cells, where the states are distinguished by inspecting their parities and field patterns. (c) Pressure distributions of the coexisting bulk and edge states at the TBIC frequencies $f_1$ and $f_2$.

*Acoustic realization of the 1D first-order TBICs.*—The tight-binding models in Fig. 2 can be implemented with acoustic cavity-tube structures. Figure 3(a) shows our acoustic realization of the mirror-stacked bilayer SSH model, where each unit cell consists of four identical air cavities coupled with narrow tubes. Physically, the cavity resonators mimic atomic orbitals and the narrow tubes introduce hoppings between them [68-72]. The structure (see details in SM [64]) provides effectively the onsite energy $\sim 7.15$ kHz, the intralayer couplings $t_0 \approx 0.12$ kHz and $t_1 \approx 0.52$ kHz, and the interlayer coupling $t_c \approx 0.25$ kHz. The successful acoustic emulation of the tight-binding model can be seen in Fig. 3(b) (left panel), where the simulated band structure captures precisely the tight-binding one. As expected, the numerical frequency spectrum for a finite-sized sample [Fig. 3(b), right panel] exhibits clearly the coexistence of topological edge states and bulk states, as guided by the horizontal dashed lines at the mid-gap frequencies $f_1 \approx 6.87$ kHz and $f_2 \approx 7.42$ kHz. The presence of the TBICs can be directly visualized from the pressure patterns in Fig. 3(c): at $f_1$ the edge-localized state of even parity coexists with the degenerate bulk state of odd parity, while at $f_2$ the bulk state of even parity coexists with the edge-localized state of odd parity.

The first-order TBICs were identified by acoustic experiments. Figure 4(a) shows our experimental sample printed precisely by photosensitive resin material with a wall thickness of 1.8 mm. On each resonator, small holes were perforated for inserting the sound source and probe, which were sealed when not in use. Both the input and output signals were recorded and frequency-resolved with a multianalyzer system (B&K Type 3560C). To measure the bulk band structure, we placed two pointlike broadband sources in the middle of the sample and scanned the acoustic response over the sample. Specifically, following the pressure distributions in Fig. 3(c), we selectively excited the states of even and odd subspaces by inphase and antiphase excitations, respectively. Figure 4(b) presents the Fourier spectra performed for the measured time-space sound signals (see SM [64]). The data for both excitations exhibit excellent agreements with the theoretical band structures. To further identify the selectively-excited mirror-symmetric and -antisymmetric states, we extracted the normalized wavefunctions $|p(x)\rangle$ and calculated the expectation values of the mirror operator, i.e. $\mathcal{M}_z = \langle p|M_z|p\rangle$, for both excitations. The results are presented in Fig. 4(b). As expected, $\mathcal{M}_z$ approaches $+1$ ($-1$) in the case of inphase (antiphase) excitation, which points to the mirror-symmetric (-antisymmetric) state of eigenvalue $+1$ ($-1$).

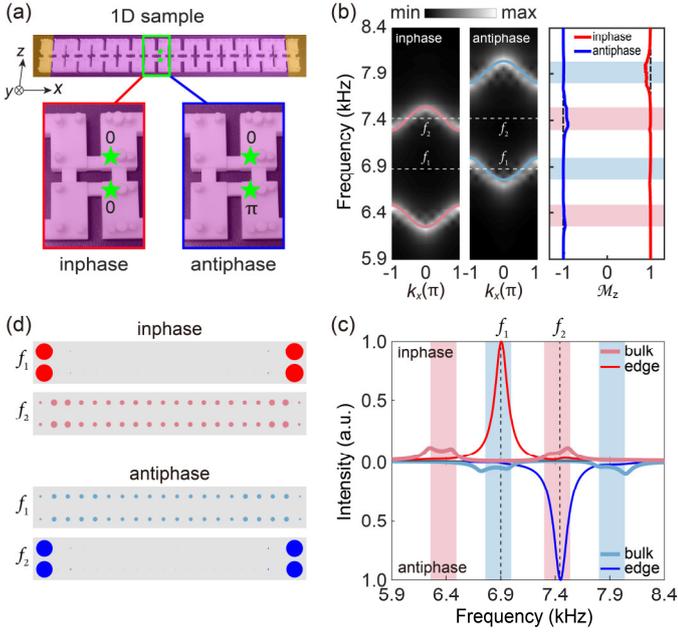

FIG. 4. Experimental evidence for the presence of TBICs in the mirror-stacked bilayer SSH system. (a) Experimental sample. It is divided into the bulk (purple) and edge (yellow) domains for extracting the data in (c). Insets: Inphase and antiphase excitations realized by a pair of sources (green stars) with 0 and $\pi$ phase shift, respectively. (b) Left: Experimentally measured band structures (bright color) with inphase and antiphase excitations, matching well the theoretically predicted even and odd bands (color lines). Right: Mirror expectation value spectra extracted from the measured wavefunctions. The color shadows indicate the frequency windows of the bulk bands. (c) Average intensity spectra measured for the bulk and edge states with different excitations. The dashed lines highlight two TBIC frequencies $f_1$ and $f_2$. (d) Intensity patterns at $f_1$ and $f_2$. The intensity is proportional to the area of the solid circle.

To directly visualize the symmetry-protected TBICs, we divided the sample into the edge and bulk regions and measured their site-resolved local responses to different excitations (see SM [64]). Figure 4(c) presents the average intensity spectra of the two regions. Clearly, the edge spectrum of the inphase (antiphase) excitation demonstrates a prominent peak at $f_1 \approx 6.90$ kHz ($f_2 \approx 7.45$ kHz), very close to that of the predicted topological edge state with even (odd) parity, i.e., $f_1 \approx 6.87$ kHz ($f_2 \approx 7.42$ kHz). As a hallmark of the TBIC, each peak falls into the frequency window of the bulk band of opposite parity. From the acoustic response, a quality factor of ~46 can be estimated for our acoustic TBIC, which is comparable to that reported for an acoustic BIC formed by symmetry protection mechanism [25]. Apparently, it can be further enhanced in a system consisting of low-loss media. To further confirm the localized nature of the topological bound states at the TBIC frequencies $f_1$ and $f_2$, we present the spatial distributions of the sound intensity at these frequencies [Fig. 4(d)]. It shows clearly that at $f_1$ the inphase excitation ignites an edge-localized state (of even parity), while the antiphase excitation induces an extended bulk state (of odd parity). The situation is reversed at $f_2$. The coexistence of the even (odd) edge state and the odd (even) bulk state at $f_1$ ($f_2$) identifies the presence of the TBIC in the mirror-stacked bilayer SSH model. Finally, we also measured the transmission responses to both excitations (see SM [64]), and provided another independent but complete evidence for the symmetry-protected TBICs.

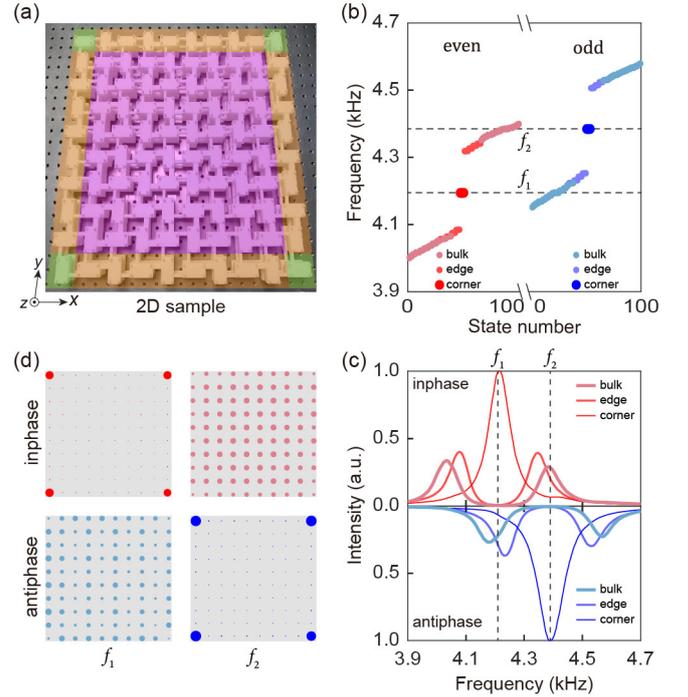

FIG. 5. Observation of the higher-order TBICs in the mirror-stacked bilayer quadrupole model. (a) Experimental sample. It is divided into the bulk (purple), edge (yellow), and corner (green) regions for extracting the data in (c). (b) Energy spectra simulated for the finite-sized system, where $f_1$ and $f_2$ denote two TBIC frequencies. (c) Average intensity spectra measured for the bulk, edge, and corner states with different excitations. (d) Intensity patterns at $f_1$ and $f_2$.

*Acoustic realization of the 2D higher-order TBICs.*—Figure 5(a) shows our experimental sample for the mirror-stacked bilayer quadrupole model. It consists of $5 \times 5$ unit cells, associated with 200 air cavities in total. Figure 5(b) provides the eigenfrequency spectra simulated for the finite-sized sample. It shows clearly the coexistence of the spectrally isolated corner states and the continuum bulk states of the opposite parity, as a numerical manifestation of the TBICs. To experimentally identify the higher-order band topology in both the even and odd subspaces, we divided the sample into the bulk, edge, and corner regions, and measured their site-resolved local responses to the inphase and antiphase excitations. Figure 5(c) presents the average intensity spectra for the three representative spatial regions, where the peaks unveil the well-excited even (top panel) and odd (bottom panel) states of the corresponding spatial domains. In particular, the inphase (antiphase) corner spectrum shows a predominant peak at $f_1 \approx 4.22$ kHz ($f_2 \approx 4.39$ kHz), which falls inside the frequency window of the odd (even) bulk band, as a spectrum hallmark for the presence of the higher-order TBIC. To further characterize the higher-order TBICs, we present the spatial intensity distributions at the frequencies $f_1$ and $f_2$ [Fig. 5(d)]. It shows unambiguously that at $f_1$ the inphase excitation ignites a corner-localized bound state while the antiphase excitation induces an extended bulk state. Similar phenomena can be observed at $f_2$, but with bulk and corner states responded to the inphase and antiphase excitations, respectively.

*Conclusions.*—We have theoretically proposed and experimentally demonstrated a simple mirror-stacking approach for constructing symmetry-protected TBICs. Interestingly, the bilayer Hamiltonian can be decomposed into two mirror subspaces of opposite parities, each of which simply inherits the band topology and the energy spectrum information (up to a shifted onsite energy) of the original monolayer. This enables a direct prediction for the presence of TBICs. Besides, our mechanism for achieving BICs can apply to an arbitrary monolayer model with bound states, even regardless their topological essence and spatial locations (see SM [64]). Both advantages benefit from the independence of the symmetries that protect the band topology and BICs. Moreover, our findings can be easily extended to other classical wave systems, which pave the way for further studies on the symmetry-protected TBICs and associated promising applications.

**Acknowledgements**
This project is supported by the National Natural Science Foundation of China (Grant No. 11890701 and 12104346), and the Young Top-Notch Talent for Ten Thousand Talent Program (2019-2022).

**References**

[1] J. von Neumann, and E. Wigner, Über merkwürdige diskrete Eigenwerte. Phys. Z. **30**, 465–467 (in German) (1929).
[2] C. W. Hsu, B. Zhen, A. D. Stone, J. D. Joannopoulos, and M. Soljačić, Bound states in the continuum, Nat. Rev. Mater. **1**, 16048 (2016).
[3] E. N. Bulgakov, and A. F. Sadreev, Bound states in the continuum in photonic waveguides inspired by defects, Phys. Rev. B **78**, 075105 (2008).
[4] D. C. Marinica, A. G. Borisov, and S. V. Shabanov, Bound States in the Continuum in Photonics, Phys. Rev. Lett. **100**, 183902 (2008).
[5] Y. Plotnik, O. Peleg, F. Dreisow, M. Heinrich, S. Nolte, A. Szameit, and M. Segev, Experimental Observation of Optical Bound States in the Continuum, Phys. Rev. Lett. **107**, 183901 (2011).
[6] C. W. Hsu, B. Zhen, J. Lee, S. L. Chua, S. G. Johnson, J. D. Joannopoulos, and M. Soljacic, Observation of trapped light within the radiation continuum, Nature **499**, 188 (2013).
[7] S. Weimann, Y. Xu, R. Keil, A. E. Miroshnichenko, A. Tunnermann, S. Nolte, A. A. Sukhorukov, A. Szameit, and Y. S. Kivshar, Compact surface Fano states embedded in the continuum of waveguide arrays, Phys. Rev. Lett. **111**, 240403 (2013).
[8] Y. Yang, C. Peng, Y. Liang, Z. Li, and S. Noda, Analytical perspective for bound states in the continuum in photonic crystal slabs, Phys. Rev. Lett. **113**, 037401 (2014).
[9] B. Zhen, C. W. Hsu, L. Lu, A. D. Stone, and M. Soljacic, Topological nature of optical bound states in the continuum, Phys. Rev. Lett. **113**, 257401 (2014).
[10] S. Mukherjee, A. Spracklen, D. Choudhury, N. Goldman, P. Ohberg, E. Andersson, and R. R. Thomson, Observation of a Localized Flat-Band State in a Photonic Lieb Lattice, Phys. Rev. Lett. **114**, 245504 (2015).
[11] R. A. Vicencio, C. Cantillano, L. Morales-Inostroza, B. Real, C. Mejia-Cortes, S. Weimann, A. Szameit, and M. I. Molina, Observation of Localized States in Lieb Photonic Lattices, Phys. Rev. Lett. **114**, 245503 (2015).
[12] J. Gomis-Bresco, D. Artigas, and L. Torner, Anisotropy-induced photonic bound states in the continuum, Nat. Photon. **11**, 232 (2017).
[13] A. Kodigala, T. Lepetit, Q. Gu, B. Bahari, Y. Fainman, and B. Kante, Lasing action from photonic bound states in continuum, Nature **541**, 196 (2017).
[14] S. I. Azzam, V. M. Shalaev, A. Boltasseva, and A. V. Kildishev, Formation of Bound States in the Continuum in Hybrid Plasmonic-Photonic Systems, Phys. Rev. Lett. **121**, 253901 (2018).
[15] A. Cerjan, C. W. Hsu, and M. C. Rechtsman, Bound States in the Continuum through Environmental Design, Phys. Rev. Lett. **123**, 023902 (2019).
[16] Z. Yu, X. Xi, J. Ma, H. K. Tsang, C.-L. Zou, and X. Sun, Photonic integrated circuits with bound states in the continuum, Optica **6**, 1342 (2019).
[17] S. I. Azzam and A. V. Kildishev, Photonic Bound States in the Continuum: From Basics to Applications, Adv. Opt. Mater. **9**, 2001469 (2021).
[18] M. S. Hwang, H. C. Lee, K. H. Kim, K. Y. Jeong, S. H. Kwon, K. Koshelev, Y. Kivshar, and H. G. Park, Ultralow-threshold laser using super-bound states in the continuum, Nat. Commun. **12**, 4135 (2021).
[19] M. Kang, S. Zhang, M. Xiao, and H. Xu, Merging Bound States in the Continuum at Off-High Symmetry Points, Phys. Rev. Lett. **126**, 117402 (2021).
[20] S. Vaidya, W. A. Benalcazar, A. Cerjan, and M. C. Rechtsman, Point-Defect-Localized Bound States in the Continuum in Photonic Crystals and Structured Fibers, Phys. Rev. Lett. **127**, 023605 (2021).
[21] Q. Zhou, Y. Fu, L. Huang, Q. Wu, A. Miroshnichenko, L. Gao, and Y. Xu, Geometry symmetry-free and higher-order optical bound states in the continuum, Nat. Commun. **12**, 4390 (2021).
[22] L. Huang, W. Zhang, and X. Zhang, Moiré Quasibound States in the Continuum, Phys. Rev. Lett. **128**, 253901 (2022).
[23] M. Kang, Z. Zhang, T. Wu, X. Zhang, Q. Xu, A. Krasnok, J. Han, and A. Alu, Coherent full polarization control based on bound states in the continuum, Nat. Commun. **13**, 4536 (2022).
[24] R. Parker, Resonance effects in wake shedding from parallel plates: some experimental observations, J. Sound Vib. **4**, 62 (1966).
[25] N. A. Cumpsty, and D. S. Whitehead, The excitation of acoustic resonances by vortex shedding, J. Sound Vib. **18**, 353 (1971).
[26] D. V. Evans, M. Levitin, and D. Vassiliev, Existence theorems for trapped modes, J. Fluid Mech. **261**, 21 (1994).
[27] C. M. Linton and P. McIver, Embedded trapped modes in water waves and acoustics, Wave Motion **45**, 16 (2007).

[28] A. A. Lyapina, D. N. Maksimov, A. S. Pilipchuk, and A. F. Sadreev, Bound states in the continuum in open acoustic resonators, J. Fluid Mech. **780**, 370 (2015).

[29] L. Huang *et al.*, Sound trapping in an open resonator, Nat. Commun. **12**, 4819 (2021).

[30] I. Deriy, I. Toftul, M. Petrov, and A. Bogdanov, Bound States in the Continuum in Compact Acoustic Resonators, Phys. Rev. Lett. **128**, 084301 (2022).

[31] S. Huang *et al.*, Extreme sound confinement from quasi bound states in the continuum, Phys. Rev. Appl. **14**, 21001 (2020).

[32] M. McIver, An example of non-uniqueness in the two-dimensional linear water wave problem, J. Fluid Mech. **315**, 257 (1996).

[33] L. S. Cederbaum, R. S. Friedman, V. M. Ryaboy, and N. Moiseyev, Conical intersections and bound molecular states embedded in the continuum, Phys. Rev. Lett. **90**, 013001 (2003).

[34] M. Robnik, A simple separable Hamiltonian having bound states in the continuum, J. Phys. A **19**, 3845 (1986).

[35] J. U. Nockel, Resonances in quantum-dot transport, Phys. Rev. B **46**, 15348 (1992).

[36] N. Rivera, C. W. Hsu, B. Zhen, H. Buljan, J. D. Joannopoulos, and M. Soljacic, Controlling Directionality and Dimensionality of Radiation by Perturbing Separable Bound States in the Continuum, Sci. Rep. **6**, 33394 (2016).

[37] H. Friedrich and D. Wintgen, Interfering resonances and bound states in the continuum, Phys. Rev. A **32**, 3231 (1985).

[38] F. H. Stillinger and D. R. Herrick, Bound states in the continuum, Phys. Rev. A **11**, 446 (1975).

[39] M. I. Molina, A. E. Miroshnichenko, and Y. S. Kivshar, Surface bound states in the continuum, Phys. Rev. Lett. **108**, 070401 (2012).

[40] G. Corrielli, G. Della Valle, A. Crespi, R. Osellame, and S. Longhi, Observation of surface states with algebraic localization, Phys. Rev. Lett. **111**, 220403 (2013).

[41] J. H. Yang *et al.*, Low-Threshold Bound State in the Continuum Lasers in Hybrid Lattice Resonance Metasurfaces, Laser Photonics Rev. **15**, 2100118 (2021).

[42] M.-S. Hwang, K.-Y. Jeong, J.-P. So, K.-H. Kim, and H.-G. Park, Nanophotonic nonlinear and laser devices exploiting bound states in the continuum, Commun Phys. **5**, 106 (2022).

[43] S. Mohamed, J. Wang, H. Rekola, J. Heikkinen, B. Asamoah, L. Shi, and T. K. Hakala, Controlling Topology and Polarization State of Lasing Photonic Bound States in Continuum, Laser Photonics Rev. **16**, 2100574 (2022).

[44] M. Z. Hasan and C. L. Kane, Colloquium: Topological insulators, Rev. Mod. Phys. **82**, 3045 (2010).

[45] J. E. Moore, The birth of topological insulators, Nature **464**, 194 (2010).

[46] X.-L. Qi and S.-C. Zhang, Topological insulators and superconductors, Rev. Mod. Phys. **83**, 1057 (2011).

[47] L. Lu, J. D. Joannopoulos, and M. Soljačić, Topological photonics, Nat. Photonics **8**, 821 (2014).

[48] T. Ozawa et al., Topological photonics, Rev. Mod. Phys. **91**, 015006 (2019).

[49] G. Ma, M. Xiao, and C. T. Chan, Topological phases in acoustic and mechanical systems, Nat. Rev. Phys. **1**, 281 (2019).

[50] B. Xie, H.-X. Wang, X. Zhang, P. Zhan, J.-H. Jiang, M. Lu, and Y. Chen, Higher-order band topology, Nat. Rev. Phys. **3**, 520 (2021).

[51] S. D. Huber, Topological mechanics, Nat. Phys. **12**, 621 (2016).

[52] B. J. Yang, M. Saeed Bahramy, and N. Nagaosa, Topological protection of bound states against the hybridization, Nat. Commun **4**, 1524 (2013).

[53] Y. X. Xiao, G. Ma, Z. Q. Zhang, and C. T. Chan, Topological Subspace-Induced Bound State in the Continuum, Phys. Rev. Lett. **118**, 166803 (2017).

[54] Z.-G. Chen, C. Xu, R. Al Jahdali, J. Mei, and Y. Wu, Corner states in a second-order acoustic topological insulator as bound states in the continuum, Phys. Rev. B **100**, 075120 (2019).

[55] X. Ni, M. Weiner, A. Alu, and A. B. Khanikaev, Observation of higher-order topological acoustic states protected by generalized chiral symmetry, Nat. Mater. **18**, 113 (2019).

[56] M. Takeichi and S. Murakami, Topological linelike bound states in the continuum, Phys. Rev. B **99**, 035128 (2019).

[57] W. A. Benalcazar and A. Cerjan, Bound states in the continuum of higher-order topological insulators, Phys. Rev. B **101**, 213901 (2020).

[58] A. Cerjan, M. Jurgensen, W. A. Benalcazar, S. Mukherjee, and M. C. Rechtsman, Observation of a Higher-Order Topological Bound State in the Continuum, Phys. Rev. Lett. **125**, 213901 (2020).

[59] Z. Li, J. Wu, X. Huang, J. Lu, F. Li, W. Deng, and Z. Liu, Bound state in the continuum in topological inductor–capacitor circuit, Appl. Phys. Lett. **116**, 263501 (2020).

[60] Z. Zhang, Z. Lan, Y. Xie, M. L. N. Chen, W. E. I. Sha, and Y. Xu, Bound Topological Edge State in the Continuum for All-Dielectric Photonic Crystals, Phys. Rev. Appl. **16**, 064036 (2021).

[61] Z. Hu *et al.*, Nonlinear control of photonic higher-order topological bound states in the continuum, Light Sci. Appl. **10**, 164 (2021).

[62] Y. Wang *et al.*, Quantum superposition demonstrated higher-order topological bound states in the continuum, Light Sci. Appl. **10**, 173 (2021).

[63] The similarity transformation matrix, which is common and explicit in our mirror-stacking approach, makes our scheme easier to generalize. This is much different from that involved in Ref. [53], which is model dependent and strongly relies on insight.

[64] See Supplemental Material at <url> for more theoretical, numerical, and experimental details, which includes Ref. [65].

[65] Y. Deng, W. A. Benalcazar, Z. G. Chen, M. Oudich, G. Ma, and Y. Jing, Observation of Degenerate Zero-Energy Topological States at Disclinations in an Acoustic Lattice, Phys. Rev. Lett. **128**, 174301 (2022).

[66] W. A. Benalcazar, B. A. Bernevig, and T. L. Hughes, Quantized electric multipole insulators, Science **357**, 61 (2017).

[67] W. A. Benalcazar, B. A. Bernevig, and T. L. Hughes, Electric multipole moments, topological multipole moment pumping, and chiral hinge states in crystalline insulators, Phys. Rev. B **96**, 245115 (2017).

[68] K. H. Matlack, M. Serra-Garcia, A. Palermo, S. D. Huber, and C. Daraio, Designing perturbative metamaterials from discrete models, Nat. Mater. **17**, 323 (2018).

[69] X. Ni, M. Li, M. Weiner, A. Alu, and A. B. Khanikaev, Demonstration of a quantized acoustic octupole topological insulator, Nat. Commun. **11**, 2108 (2020).

[70] H. Xue, Y. Ge, H. X. Sun, Q. Wang, D. Jia, Y. J. Guan, S. Q. Yuan, Y. Chong, and B. Zhang, Observation of an acoustic octupole topological insulator, Nat. Commun. **11**, 2442 (2020).

[71] Y. Qi, C. Qiu, M. Xiao, H. He, M. Ke, and Z. Liu, Acoustic Realization of Quadrupole Topological Insulators, Phys. Rev. Lett. **124**, 206601 (2020).

[72] J. Du, T. Li, X. Fan, Q. Zhang, and C. Qiu, Acoustic Realization of Surface-Obstructed Topological Insulators, Phys. Rev. Lett. **128**, 224301 (2022).